# Underground Astronauts: Understanding the Sporting Science of Speleology and its Implications for HCI

Authors: Eleonora Mencarini[1*], Amon Rapp[2], Massimo Zancanaro[3&1]

[1]Intelligent Interfaces and Interaction research unit (i3), Fondazione Bruno Kessler (FBK), Trento, Italy

[2]Department of Computer Science, University of Torino, Torino, Italy

[3]Dept. of Cognitive Science, University of Trento, Rovereto (TN), Italy

*corresponding author: mencarini@fbk.eu

## Abstract

In this paper, we present a qualitative study on speleology that aims to widen the current understanding of people's practices in Nature and identify a design space for technology that supports such practices. Speleology is a practice based on the discovery, study, and dissemination of natural cavities. Speleologists are amateur experts who often collaborate with scientists and local institutions to understand the geology, hydrology, and biology of a territory. Their skills are at the same time physical, technical, and theoretical; this is why speleology is defined as a 'sporting science'. Being at the boundary between outdoor adventure sports and citizen science, speleology is an interesting case study for investigating the variety and complexity of activities carried out in the natural context. We interviewed 15 experienced speleologists to explore their goals, routines, vision of the outdoors, and attitude towards technology. From our study, it emerged that i) the excitement of discovery and the unpredictability of an explorative trip are the strongest motivations for people to engage in speleology; ii) physical skilfulness is a means for knowledge generation; iii) the practice is necessarily collective and requires group coordination. From these findings, an ambivalent attitude towards technology emerged: on the one hand, the scientific vocation of speleology welcomes technology supporting the development of knowledge; on the other hand, aspects typical of adventure sports lead to resistance to technology facilitating the physical performance. We conclude the article by presenting design considerations for devices supporting speleology, as well as a few reflections on how communities of speleologists can inspire citizen science projects.

## 1. Introduction

Over the recent years, HCI has investigated how technology can support human activities related to Nature (Bidwell and Browning, 2010; Häkkilä et al., 2018; McCrickard et al., 2020; Su and Cheon, 2017), such as learning (Fails et al., 2014), outdoor sports (Anderson et al., 2017; Cheverst et al., 2020; Tholander and Nylander, 2015; Woźniak et al., 2017), adventure (Müller and Pell, 2016), and 'citizen science', an activity where citizens are involved in data collections for environmental sciences projects (Cottman-Fields et al., 2013; Moran et al., 2014; Phillips et al., 2014; Tinati et al., 2015). In these practices, Nature is typically conceived as a place where to find restoration, silence, and liberation from routines (Häkkilä et al., 2018). There, people enjoy themselves by challenging natural elements or studying natural phenomena (Davidson and Stebbins, 2011). However, HCI studies have investigated these different ways to experience the outdoors by univocally ascribing them to either the sports domain or the knowledge generation and dissemination domain. These works have highlighted the challenges that technology design faces when aimed at supporting human practices in Nature (Cheverst et al., 2020; Daiber et al., 2016; Häkkilä et al., 2018; Mencarini et al., 2019) and showed very different attitudes towards technology by people engaged in one domain or the other. For example, while hikers, climbers, and skiers show some resistance to using technology in the wild, pushing HCI to reflect on its role in natural environments (Cheverst et al., 2020; Häkkilä et al., 2018; Mencarini et al., 2019), citizen scientists are usually open to technological artefacts for collecting and sharing information (Gaver et al., 2019; Hsu et al., 2017; Phillips et al., 2014), seeing them as a way to democratise science.

By contrast, in this article, we focus on a practice that merges the aspects of both domains: speleology. Speleology is the exploration of caves for scientific purposes; it has the ultimate goal of discovering new caves, studying them, and sharing this new knowledge with others. It is often defined as a 'sporting science' because it blends sports and science, similarly to snow science, scientific diving, and wildlife sightings. Since it involves user needs that pertain to both scientific and adventurous explorations, we believe that speleology may widen our understanding of how technology may support activities related to Nature as well as offer insights on the sustainable functioning of citizen science projects.

In this work, we adopt the practice paradigm conceived by Kuutti and Bannon (2014), in which the interaction with technology is not considered as a dyadic relationship between the user and the technological artefact but takes place in a broader context made of human relations, communities, shared activities, and artefacts. In particular, we frame our study within the realm of *peripheral practices*, namely "niche, unusual, marginalized and/or highly specialized communities of practice, [whose study] results in implications for HCI outside that community" (Tanenbaum and Tanenbaum, 2018, p.11). Investigating peripheral practices is interesting because they introduce a diverse perspective on existing problems, serving as a defamiliarizing lens. By adopting this lens, we aim to both *design for speleology*, that is, to define suggestions for creating technologies specifically addressed to speleologists, and *learn from speleology*, that is, to derive insights from speleology to support citizen science.

To understand speleology, we investigated: i) speleologists' culture and values; ii) their organisation and routines, and iii) their use of existing technology and openness to new digital tools. Our findings highlight that speleology is primarily a physical practice that happens outdoors, in which, nonetheless, physical effort and self-challenge are subordinate to higher goals, such as the discovery of unexplored places and the gaining of scientific knowledge. Speleologists show interest in tools - both digital and non-digital - that can help them perform their scientific activity. Typically, these tools are either appropriated from other domains (e.g., architecture, backcountry skiing, and plumbing) or self-built. However, they are sceptical about technology that may replace or greatly ease the physical performance required to explore the caves, highlighting the sports nature of speleology. In sum, this study unveils speleologists' need for technological devices that support their desire for knowing, measuring, and documenting, but that do not jeopardise their physical conquest of Nature.

This work contributes to both HCI for outdoor adventure sports and HCI for citizen science. As to the former, it provides insights for the design of technologies that merge sports and documentation for scientific purposes; as to the latter, it gives suggestions on how citizen science projects can be structured by leveraging people's competencies and interests.

The article is structured as follows: first, we introduce the general characteristics of speleology. Second, we discuss the related work in the field of outdoor adventure sports and citizen science. Then, we present the interviews we conducted with expert cavers and the main themes describing how the practice of speleology is articulated. Finally, we present a series of considerations for designing technology for speleology, as well as a series of lessons that citizen science could learn from grassroots practices.

## 2. Speleology

Speleology was identified for the first time as a 'sporting science' at the end of the 19th century by Édouard-Alfred Martel, one of the key promoters of this discipline (Cant, 2006; Pérez, 2015). This two-fold nature of speleology can already be found in its definition: according to the Cambridge Dictionary, the word *speleology* means both i) the scientific study of caves and ii) the sport of walking and climbing caves[1]. Synonyms are present as well: *caving* (UK) and *spelunking* (US); however, these latter terms lean more towards the sports interpretation. Sports and science are tightly intertwined in speleology because, on the one hand, speleologists differ from cavers and miners for their scientific purpose (Mattes, 2015); and on the other hand, their scientific goal could not be achieved without a great physical engagement. Thus, developing speleological knowledge depends on exploration, and the greater the physical abilities of the explorers, the greater the likelihood that more caves will be discovered and surveyed (Pérez, 2015).

Indeed, cave exploration is very demanding from the physical point of view. Caves can unfold both horizontally and vertically. Some horizontal passages can be walked to pass through while others require to squeeze flat (Figure 1b), some ledges can be slippery or on top of pits, and vertical passages require strength to be ascended because the progression occurs along a rope by pushing the whole-body weight up through a stirrup (Figure 1c). Furthermore, speleologists are loaded with a lot of equipment: ropes, hammers, drills, batteries, fixes, water, food, emergency kits, survey tools, cameras, etc. Because of these difficulties, speleology is a collective endeavour, where effective group coordination and efficient teamwork are extremely important.

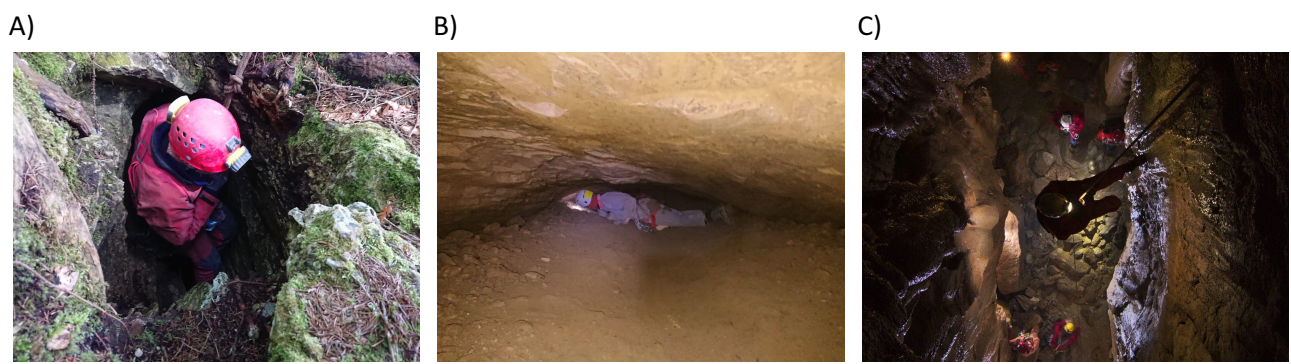

**Figure 1. a) Entering a cave, b) Squeeze flat – © GST; c) Climbing back from the bottom of a cave © GSL.**

However, the practice of speleology consists not only in exploration, but also in a preliminary phase of discovery, through the study of documentation and the observation of the territory, and a subsequent phase of survey and dissemination of knowledge to the local authorities, other speleologists, academics, and the general population. Dissemination is taken seriously by speleologists since "caves are hidden places; we learn of them as cavers return to the surface, translating and narrating their explorations vividly through stories and images" (Cant, 2003, p.68). Although the practical science of speleology contributes to several academic disciplines such as geography, physical, geology, hydrology, biology, meteorology, ecology, archaeology, anthropology, etc. (Mattes, 2015), there are few departments of Speleology in universities around the world and the activity is mostly carried out by normal people (Pérez, 2015). For this reason, in most cases speleology can be considered as a form of voluntary and self-organized citizen science practice.

---

[1] SPELEOLOGY | entry in the Cambridge English Dictionary,
https://dictionary.cambridge.org/dictionary/english/speleology, accessed 27/03/2020.

## 3. Related work

In this section, we first discuss previous literature on outdoor adventure sports since speleology is an activity that happens outdoors, requires a physical performance, and is often adventurous. Then, we present an overview of studies in the domain of citizen science. Even if speleology is not a 'proper' citizen science project, but a self-organised community of practice carried out by expert amateurs and aimed at generating knowledge about the environment, we believe that comparing speleology and citizen science could be mutually informative. On the one hand, citizen science provides us with a lens for understanding the motivations, practices, and goals of speleologists; on the other hand, the organisation of speleology groups might provide useful insights to address the citizen science issues identified by the literature. Finally, we conclude the section by presenting previous works exploring technological applications for speleology.

**3.1. Outdoor adventure sports**

Although there is no clear classification of outdoor adventure sports, they are typically defined as activities happening in a wild environment involving risk, challenge, and uncertain outcomes (Müller and Pell, 2016; Pike and Beames, 2013; Wheaton, 2004). The performance of these sports consists in the tackling of a natural element, such as the verticality of mountain walls in climbing, the deepness of water in scuba-diving, the unpredictability of snow in backcountry skiing, and requires physical preparation, knowledge of the environment, specific gear to face it, psychological and emotional firmness, and coordination with partners, as they are usually conducted in a group to minimise risk.

The topic of adventure has not been extensively investigated in HCI yet. A remarkable exception is Müller and Pell (2016), who articulated four possible roles that technology can take in remote environments according to the possible expected or unexpected events that can happen. These are the *coach* role, guiding in for a better performance during expected situations; the *rescuer* role providing emergency services during unexpected events, as avalanche beacons do; the *documentarian* role for recording expected events with a high experiential value; the *mentor* role supporting the adventurer's reflection on the felt experience during unexpected events.

Conversely, the topic of technology for outdoor sports has received more attention in HCI. A few works have focused on the acceptance or resistance to technology by outdoor sportspeople during their practices. Cheverst et al. (2020) highlighted the importance of the feeling of mastery in mountaineering, as it motivates practitioners and helps them comply with the rules of their community. In such a context, technologies easing environmental challenges and reducing the required technical skills are perceived as cheating and thus strongly discouraged, while technologies aimed at recording the performance are welcomed. Similarly, Ahtinen et al. (2008) noticed that applications in the form of a logger or personal diary for tracking physical performance would be welcomed because they allow outdoor sportspeople to challenge themselves over time. The will to gain all the necessary competencies to face an outdoor sports adventure was also at the root of the reluctance that Mencarini et al. (2019) found when introducing wearable devices in climbing. They discovered that climbers fear that these kinds of devices might contradict their core values, i.e., self-efficacy, mutual trust between partners, and the ability to manage the unexpected events that an adventure may bring along.

Given the widespread outdoor sportspeople's resistance to technology 'mediating' their challenge with natural elements, HCI focused on either supporting unmediated outdoor experiences or providing essential information to enable the activity. As for the first group, Tholander and Nylander (2015) noticed that outdoor sportspeople enjoy to some extent the annoying feelings of pain, sweat, and fatigue that these sports entail and suggest considering those sensations and the overall sportspeople's personal lived experience in the design of technology; likewise, Posti et al. (2014) developed an app that ensures solitude while hiking by suggesting solitary paths. As for the second group, HCI works have investigated the importance of information sharing in terms of trustworthiness of the source, safety, and privacy (Daiber et al., 2017; Woźniak et al., 2017), as well as the effectiveness of technology designed to provide support in situations of emergency, where collaboration and promptness are fundamental (Desjardins et al., 2014).

To summarise, these studies highlight that people engaging in outdoor adventure sports seek direct contact with natural elements, adventure uncertainty, and challenge their physical and mental abilities. Therefore, outdoor adventure sportspeople are not inclined to use technology aimed at facilitating or mediating their experience; if at all, they are interested in keeping track of that experience. Things change in situations of danger when people want to be in full control. In such cases, a prompt technology providing the right information is a safeguard against risk. We build upon the literature presented here to highlight the tension between adventure, mastery, and control. Still, we acknowledge a gap in the study of outdoor practices that pursue various goals, such as those of sporting sciences, which we address in this work.

**3.2 Citizens Science**

The term 'Citizen Science' indicates projects with a scientific aim based on the collaboration of citizens to collect, analyse, or disseminate data. The citizens involved in these projects are long-term groups of volunteers with variable levels of skills and domain knowledge. This way of doing science is believed to bring multiple advantages: the scientific community benefits from the collection of data that would not have been possible to collect otherwise (or at least not in such a large quantity); citizens can be educated about science; and, finally, collective attention can be drawn on societal concerns, like the quality of life and ecosystem balance (Aoki et al., 2017).

So far, research in HCI on citizen science explored i) the different levels of citizens' engagement, which denote specific goals and types of citizen science projects (Aoki et al., 2017; Bonney et al., 2009; Qaurooni et al., 2016); ii) the sustainability of citizen science projects in the long term (Mugar et al., 2014; Rotman et al., 2012; Tinati et al., 2015); iii) the quality of data based on the level of citizens' expertise (Elbroch et al., 2011); iv) and the design of technology supporting the activity, that is, to collect, upload, analyse, or share data (Cottman-Fields et al., 2013; Gaver et al., 2019; Phillips et al., 2014; Polys et al., 2020). Below, we present the main issues addressed in the literature regarding each of these points.

Citizen science projects can be seen as a continuum spanning from the least engaging to the most engaging for citizens who take part in them (Bonney et al., 2009). Typically, less-engaging projects are driven by professional scientists and have the unique goal of increasing scientific knowledge. These types of projects are often defined as 'crowd science' (Qaurooni et al., 2016) or 'crowdsensing' (Aoki et al., 2017) because citizens provide only data, and their work is characterised by extreme division of labour and standardisation, while the project goal is established by professional researchers (Aoki et al., 2017). At the other end of the continuum, there are projects co-designed by scientists and people together, which aim to affect the local environment where they take place. There, participation is a way for civic engagement (Qaurooni et al., 2016). In the middle of the continuum, there are collaborative projects in which local experts or amateurs refine the project design, analyse data, and disseminate findings.

Citizen science projects come with inherently complex activities spread over long periods and spanning multiple tasks. The sustainability of these projects lies in their ability to attract and train newcomers (Mugar et al., 2014), keep participants' motivation alive in the long term (Aoki et al., 2017; Rotman et al., 2012), and accomplish the set scientific goals (Tinati et al., 2015). Because citizens' level of expertise is closely related to the quality of the data collected and, at times, analysed (Elbroch et al., 2011), training newcomers is fundamental. Mugar et al. (2014) investigated the problem of training newcomers in online citizen science projects and found a possible solution in what they call the 'practice proxies', i.e., traces of participation in online environments acting as resources to orient newcomers towards the norms of the practice. The practice proxies work as a resource for newcomers allowing them to understand how to look at the data they encounter on the shared online platform. Moreover, the groups of volunteers need to be continuously enriched and motivated. According to Aoki et al. (2017) and Rotman et al. (2012), volunteers' motivation dynamically changes throughout the project life cycle, usually starting from more egoistic reasons and then moving towards community involvement and environmental advocacy.

HCI also aimed to develop technology that democratises scientific knowledge and supports citizens' empowerment (Hsu et al., 2017). For instance, the BBC and the Goldsmiths university collaborated to give anyone interested the opportunity to build his/her own camera to take pictures of Nature (Gaver et al., 2019). Similarly, Phillips et al. (2014) wanted to enable citizens to build their own device for collecting data about bees, and Polys et al. (2020) designed an app for collecting information about the distribution and growing conditions of medicinal forest plants. Nevertheless, the introduction of technology for facilitating the study of Nature is not always unproblematic: the will to widen the volunteer base and the potential damage that this democratisation can bring to the safeguard of a habitat (which is the primary goal of many citizen science projects) may generate tensions in already established communities of amateurs (Moran et al., 2014).

### 3.3. Technology for caves

We found only five papers dealing with technology and caves, and none of them presents technology for improving or facilitating the activity of caving; rather, they focus on technology for documenting caves. One of the most popular approaches for caves documentation is to capture the internal parts of the cave through a laser scanner (Silvestre et al., 2013) and then reproduce them through 3D virtual representations, so to offer an immersive experience to tourists (Beraldin et al., 2006), students (Adcock et al., 2015), and experts of prehistorical art paintings (Wang et al., 2010). Among these five papers, only the one of Schuchardt et al. (2007) presents an HCI study where 24 cavers assessed the effectiveness of Immersive Virtual Environments in providing a spatial understanding of the complex structures of caves. The commitment to dissemination that these studies highlight also emerged from our study.

## 4. Study

This study is part of a project aimed at identifying design opportunities for wearable technology in outdoor adventure sports.

**4.1. Goal**

The purpose of this study was to explore the practices of speleology, better understand its twofold nature of 'sporting science' and identify possible suggestions for further developing HCI literature at the boundary between sports and citizen science.

**4.2. Methodology**

We conducted participated observations during the lessons of two introductory courses on speleology organized by two caving groups in the Trentino province (Italy), which involved 16 beginners and 14 instructors in total. Then, we recruited 15 experienced speleologists belonging to these groups (3F, 12M; average age 40 years, Standard Dev.= 14.3. A detailed demographics of our informants is illustrated in Table 1) and conducted semi-structured interviews with them. The GSL[2] group was established in 1974 and was affiliated to the Società Speleologica Italiana, whereas the GST[3] group was established in 2000 and affiliated to the Scuola Nazionale di Speleologia of the Italian Alpine Club, the two main Italian national speleological associations. Both groups were contacted by email in December 2018, and in January 2019 we had the first meeting to discuss the terms of the research and their availability. The observations and the interviews were then conducted in Spring 2019.

Table 1. Participants' demographic data.

| ID | Group | Sex | Age | Caver since | Instructor | Since | Member of the Rescue Service | Since |
|---|---|---|---|---|---|---|---|---|
| I01 | B | F | 44 | 2000 | ✓ | 2007 | - | - |
| I02 | B | M | 64 | 1972 | ✓ | 1991 | Not anymore | - |
| I03 | A | M | 55 | 1987 | - | - | - | - |
| I04 | A | M | 43 | 1994 | ✓ | 2014 | Not anymore | - |
| I05 | A | M | 33 | 2015 | ✓ | 2017 | ✓ | 2016 |
| I06 | A | F | 31 | 2015 | ✓ | 2017 | - | - |
| I07 | A | M | 60 | 1976 | - | - | - | - |
| I08 | A | M | 25 | 2010 | ✓ | 2017 | ✓ | 2013 |
| I09 | A | M | 24 | 2011 | - | - | ✓ | 2017 |
| I10 | B | M | 46 | 2006 | ✓ | 2013 | - | - |
| I11 | A | M | 26 | 2010 | ✓ | 2017 | ✓ | 2013 |
| I12 | A | M | 28 | 2012 | ✓ | 2017 | ✓ | 2016 |
| I13 | B | M | 35 | 2012 | ✓ | 2014 | ✓ | 2016 |
| I14 | B | F | 28 | 2007 | ✓ | 2013 | - | - |
| I15 | B | M | 62 | 2003 | ✓ | 2013 | - | - |

None of the authors had previous or direct experience of speleology; even if the first author has intermediate knowledge of several mountain disciplines, she has no competence in speleology. In this regard, the theoretical lessons were observed to gain an initial understanding of the discipline; whereas, interviews were conducted asking questions about past experiences, group habits and routines, personal motivations to practise speleology, and vision of the use of technology in speleology (the structure of the interview is reported in Table 2). The interviews were conducted either at the headquarters of each speleological group or bars during after-work hours. They lasted about 1 hour and 15 minutes on average and were then transcribed and analysed following the Thematic Analysis methodology (Braun and Clarke, 2006).

---

[2] Gruppo Speleologico Lavis
[3] Gruppo Speleologico Trentino

Table 2. Semi-structured interview protocol

| Topic investigated | Questions |
|---|---|
| Level of familiarity with speleology | <ul><li>How did you start practicing speleology?</li><li>Do you practice any other sport? What differentiate speleology from these sports?</li></ul> |
| Personal view | <ul><li>Why is speleology cool?</li><li>Do you think speleology can be considered a sport?</li></ul> |
| Field trips and risks | <ul><li>Would you say that speleology is an adventurous activity?</li><li>When do you agree to take a risk while caving?</li><li>What are the problems that can happen during a field trip? Could you recall an episode when you had problems?</li><li>What happens when somebody cannot take it any longer?</li><li>How do you communicate one with the other when you are in a cave?</li></ul> |
| Training new speleologists | <ul><li>What is the goal of the speleology course?</li><li>What do people must know once the course is over?</li></ul> |
| Technology and caving | <ul><li>What do you think of technology in speleology?</li><li>Do you use any device while caving?</li><li>Could you recall an episode where technology helped you while caving?</li><li>Have you ever thought 'I'd like to have a device that…'?</li></ul> |

We followed an interpretivist stance for the analysis of qualitative data (McDonald et al., 2019), an approach that is widely used in HCI and CSCW (Olson and Kellogg, 2014). Initially, the first author read and analysed the transcriptions in search of open codes. On the one hand, she searched for the elements that could explain the nature and the practice of speleology; on the other hand, she coded descriptions and expressions that revealed relevant and unexpected aspects of the practice. After conducting the open coding, the first author developed the axial codes by grouping the open codes in key categories according to the activity model of a caving trip. To this end, the notes taken during the observation of the theoretical lessons were used to guide the interpretation. Then, the second author read the interviews, reviewed the generated codes, and discussed inconsistencies in interpretations with the first author. Inconsistencies were mainly related to the labels applied to certain concepts. Furthermore, interview transcriptions and analysis have been an ongoing process: interviews of the participants belonging to the two groups were conducted in two different moments, so the second group was used to assess the reliability of the interpretations developed during the first phases of the analysis. In total, we identified 250 open codes and 18 axial codes. The four main themes motivating and articulating the speleologists' practice emerged from the final discussion among all the authors.

**4.3 Findings**
Overall, speleologists tend to consider positively the introduction of new tools, including digital technology, that could make their activity easier: *"Whatever can favour the progression, the communication, the use of other devices or the development of other activities, I tend to see it favourably"* (I03). Indeed, they have a history of both self-made tools, such as flexible ladders to descend pits or geo-radars to check for cavities in the underground, and appropriation of devices typically used in other fields. In the following subsections, we will illustrate the themes that emerged from the analysis of the interviews. Such themes explain the dynamics of the speleological practice and the role that technology plays in the different moments of the practice. Since the darkness of caves was often reported as a constitutive aspect of the practice, we framed the themes based on the interplay between light and dark. The four themes that emerged are i) Bringing the light: exploration as the main motivation and goal; ii) Lights, camera, action! The performance inside the cave; iii) A voice in the dark: group coordination and communication; iv) Unearthing: sharing discoveries with others. A summary of the themes identified is summarised in Table 3.

Table 3. Summary of themes.

| Themes | Core activities | Dimensions | Roles of technology |
| --- | --- | --- | --- |
| Bringing the light | Land and cave exploration | Discovery | Gain access to and support the exploration of hostile places (Internet, headlamps, avalanche beacons, endoscope cameras) |
| | | Imagination | Explore the unknown, preview the yet unseen (endoscope cameras) |
| | | Astronauts on Earth: exclusivity & privilege | Trace a new path walked for the first time (Strava for the cave + Google Earth) |
| Lights, camera, action! | Self and contextual awareness | Timing & weather conditions | Provide information about personal and external environmental conditions |
| | | Physical effort & mental strength | |
| A voice in the dark | Group coordination and communication | Social adventure | Communication technologies (walkie-talkies, phone cable) |
| | | Rescue operations | |
| Unearthing | Training newcomers and safeguarding memory, documentation and dissemination | Interdisciplinarity brings different roles | A variety of sub-disciplines and tools |
| | | Official reports for institutional audiences | Technologies for surveying and collecting data (GPS, Disto™ X, TopoDroid app, drones, maps) |
| | | Experiential reports for informal audiences | Technologies and artefacts for note-keeping, group self-reflection, and social memory (diaries, videos); Technologies for immersive dissemination (VR) |

*4.3.1. Bringing the light: exploration as the main motivation and goal*

Caves are hostile environments: they are dark, cold, dump, and arduous. There is no light, and there are very few life forms; there, it is possible to experience total darkness and silence. This environment fascinates people who love exploration. This first theme shows how discovery, i.e., the possibility to see new underground environments for the first time, depends on the human ability to bring artificial light in them; how having a strong imagination is important to pursue an invisible goal with perseverance like the search for new caves; and, finally, how the high level of isolation and hostility of the environment makes cavers feel as astronauts on earth, with the subsequent feelings of exclusivity and responsibility that it entails.

**Discovery**. From the interviews, it emerged that speleologists long to be the first to see a certain underground place by bringing light to it. As a few interviewees affirmed: "*That is, bringing the light where the light has never been before, and even the walls have never seen each other*" (I08); "*[Caves are] a frontier to explore, where no human being has been before. We are the first living beings reaching there, aren't we?*

*We arrive and we light it up. In an untouched cave, we are the first to bring the light*" (I11), "*I arrive, light up the darkness, and when I pass by the darkness submerges me again*" (I01). Technology, broadly intended as tools and artefacts, is essential to gain access to such places. Speleologists have their own specific gear, including ropes, descenders, handheld and chest ascenders, bolts, and drills. Headlamps in particular have a pivotal role, and they have been subject to the most appreciated technological advancements. In fact, earlier speleologists used carbide lamps, but recently the new LED lamps have been quickly adopted. Although many of our participants stated that using carbide lamps was more romantic and had the secondary benefit of keeping speleologists warm, the power, duration, water resistance, and pollution reduction (which was due to the smoke of carbide powder) favoured the adoption of LED lamps into the practice.

**Imagination**. Speleologists' main activity is to search for new caves or further explore already known caves. This willingness to discover new places despite the hostile environment and the uncertainty of gaining positive results characterises speleology as a very 'psychological' discipline since participants seem to be mainly moved by intrinsic motivation. In the words of one of the participants:

> *"Exploration is a very abstract concept: it can be a crack from which you feel some air blowing, and it seems to widen, so you try to open it up, and finally you realise that it does not widen at all. I mean, this is the beauty of caving: all the beating you get while exploring. Then, one in a thousand makes it. [...] In the end, speleology is just this, right? Imagination" (I09).*

A very similar concept can be found in Cant (2003, p.71) when she affirms that "the top of a mountain really is the top, all there is to that specific mountain, but the end of a cave passage may not be 'it'. A cave passage may simply be blocked – by silt or fallen rocks – and with a bit of help from cavers (digging away the debris), more cave can be discovered". The invisibility of what comes next is also a great source of thrill, as expressed by I03: *"when you arrive at the edge of a pit, you throw a rock and start counting the seconds before hearing the first bounce… This is the thrill of exploration"*.

Cavers use technology for explorative purposes: to understand how a cave unfolds and thus estimate the effort required for its exploration, they appropriate technology typically used in other fields. One example is avalanche beacons, i.e., transceivers designed for and used by back-country skiers to rescue people caught in avalanches. In case a cave branch progresses up towards the surface rather than down in depth, one caver would stand outside the cave on a spot where s/he thinks the branch is close to the surface, while the other would stay inside the branch, and they would search for each other's signal. Beacons are very useful because they measure how far they are from each other, so it is possible to have an idea of how much digging work is needed. Another technology that has been appropriated by the cavers is endoscope cameras, which are typically used by plumbers. In caving, these devices are employed to explore cracks that are too narrow for cavers to go through them. By looking at the images filmed by the endoscope camera, cavers can preview what is beyond a crack and decide whether it is worth to widen it and continue the exploration in that direction.

**Astronauts on earth.** The underground world is often described as the last 'bastion' still to be explored, and speleologists often feel like astronauts or explorers of a parallel world: "*Speleology [...] makes you feel like an astronaut who sets foot on the moon* [for the first time] *or a person who sees places never seen before*" (I05)*; "Speleology allows you to be some sort of astronaut. I mean, the outside world has been seen through* *scans and explorations. When exploring caves, you might get to places where not only humans have never been, but not even the light!"* (I08).

As they adventure into 'another world', speleologists need special equipment to face the unknown, which at times is not available in the market and may lead them to create their own technology:

> *"A guy who joined our group last year is thinking to create a sort of Strava for the cave. This device would take the GPS coordinates at the cave entrance, where there is still a GPS signal, and then would count the meters you walk. This system would not be able to calculate the cave size, but maybe the plotline. Even if it won't be useful for surveying, it could be used during explorations, to then have a look at Google Earth and see how long I have walked and how far from the surface I am, which direction I am going"* (I09).

Finding new places brings a sense of exclusivity and privilege: "*The top of a mountain is a place where few people can get, but caves - when exploring - are places where nobody has arrived before*" (I11). This privilege requires an assumption of responsibility and full awareness of the adventure, meaning that control is fundamental. From this perspective, the first thing a group of speleologists needs to do before leaving for an explorative trip is to search for documentation and planning. In this respect, digital technology can become a source of information (in the form of the Internet or specific digital libraries) and a way to coordinate the group before beginning the trip.

> *"The trip starts when you plan it, hopefully at least two days in advance. Typically someone in the group tells you 'I have seen a hole in that mountain wall and I think nobody has been there yet', or s/he has been in an already known cave and (…) something (…) has captured his/her attention. After that, you take a cave survey and start to measure it to quantify the amount of equipment you will need: if it is an ascent, if you need to dig, it depends... Once you have understood what you need to do, you will know how long it will take (if one day or more), and be able to decide the team, prepare the equipment…"* (I08).

In this first theme, we highlighted the main motivations, goals, sensations, and values that are the basis of the speleologists' practice. Here, technology - when used - is a support that does not hamper the feelings of excitement, discovery, effort, and agency they experience while caving.

### *4.3.2. Lights, camera, action! The performance inside the cave*
Once inside a cave, the darkness and isolation from the outside prevent cavers from perceiving the conditions of the external world and alter their self-perception, resulting in a possible source of danger. Awareness of one's left energies is important because caves require great physical performance, especially to go back to the surface. The risk of being too tired to climb back and becoming demoralised is high. Notably, the desirable solution to prevent such a situation would not be a tool to reduce the effort, but rather to make the caver constantly self-aware of his/her psycho-physical condition. Similarly, it would be beneficial to maintain awareness of the flowing of time and the external weather conditions because their sudden changes might compromise the safety of the trip.

**Physical effort and mental strength.** Caves have many different internal shapes: some of them develop mainly horizontally, others vertically. In the latter, speleologists would reach the deepest point by descending (and then climbing back) the vertical traits using ropes. In those situations, "*the rope is the* *road*" (I11), and while the descent is quite effortless, the ascent is very demanding because it occurs by pushing the whole body up along the rope through a stirrup. The effort is even higher if the goal of the trip is to discover new cave branches. In this case, cavers have to physically work to widen cracks in the rock or remove stones that block the passage. Furthermore, often there is an 'approach time' to consider, that is when cavers need to hike to reach the cave and go back from it: "*often, we need to hike to approach the cave, so when usually hikers would go back because they have arrived at the end of the hike, there we start our trip into the cave. Therefore, we push until we are dead tired*" (I14).

Mental and emotional strength to resist for a long time in such a hostile environment is needed as well. When not connected to episodes of fear, emotional breakdowns are closely related to a loss of physical strength. So, physical and mental conditions are intertwined:

> "*I have noticed that often mental stress is a consequence of a feeling of discomfort inside the cave, or of being physically tired, or worn out, or just fed up with thinking 'why the heck am I here? I want to get out!' and when you are fed up, you slow down, you begin to climb as less as possible, and a whole chain of events starts and leads to the complete demoralisation of the person*" (I11).

During the interviews, the speleologists mentioned several times a recent invention they saw on the Internet, which would lighten the physical performance. It is an electric winch that would help cavers climb back the cave by attaching it to the rope and making it retrieve them. The idea of such a device raised conflicting opinions: on the one hand, it fascinated them; on the other hand, they were afraid that it would excessively popularise speleology bringing people who are not adequately trained inside caves.

> "*Nowadays, it is quite a manual activity, very fatiguing, physical, old-style… and that is quite positive because it makes speleology less accessible, accessible only by instructed and able people*. [There is a guy who] *has built an electric winch, but he caves only deep pits because bringing it along [is such a] hassle*" (I05).

> "*Such systems* [helping the ascent from the bottom of a cave] *would allow having more people caving, maybe people less trained, who might take speleology lightly, like 'that's ok, that gadget will pull me up'. Perhaps the beauty of speleology is that it is quite a niche sport, and I think I prefer it to stay so*" (I13).

In this respect, caving resembles more those adventure sports in which athletes test their physical abilities against Nature, avoiding any technological support that could favour the accomplishment of the performance. The contrasting attitudes toward such 'empowering technologies' may thus signal a tension between the exclusivity of the sports adventure, which should be accessible only to people able to face the endeavour, and the nature of science, which should not exclude individuals based on their physical strength, thus welcoming any tool that may relieve the fatigue of the body.

**Timing and weather conditions.** Time has great relevance for some aspects of speleology and not for others. The conditions of a cave in the different seasons of the year, as well as the weather conditions, determine when caves are accessible. At times, the time span for accessing a cave is so short that it can lead to compromises: "*In many alpine caves, you have a limited time span to access them – mainly because* [they are covered with] *snow – and even if you don't find the right number of people, you still need to go because* *the exploration takes so long... and we cannot renounce every time*" (I12). Furthermore, an estimation of the overall duration of a trip and of each its phases is required:

> "[While exploring] *when we descend, there is no downtime; we go down as fast as trains. That is one of the few moments where timing counts because the less time you spend in the approach and getting to the bottom of the cave, the more time you have to spend in the spot where you need to work, dig, rig*" (I12).

Once inside, the temporal dimension loses its significance, as the time needed for the exploration is unpredictable and the absence of light creates a suspended atmosphere. Being out of the world for an indefinite time is perceived as very relaxing by speleologists. However, often cave incidents are caused by changes in the external weather conditions, such as rainstorms or snowfalls, that affect the accessibility of the cave. In the words of I03:

> "*A problem that can occur is that unexpected bad weather arrives, part of the cave gets flooded, and the passage is hindered. However, we do not have that kind of problem here because generally caves are at high altitude. What is more likely to happen [here] when we cave (...) in winter is that weather conditions change,* [it starts to snow] *and the risk of avalanches may affect tired cavers when they get out of the cave.*"

During a cave trip, weather conditions, as well as personal ones, can change without cavers realising. Therefore, planning the duration of a trip and keeping awareness of the time passing are necessary actions for ensuring safety. At present, this problem is solved just by carefully studying cave documentation and watching the weather forecasts before starting the trip.

### *4.3.3. A voice in the dark: group coordination and communication*
Speleology is a collective practice. Its compulsory collectivistic nature is due to safety reasons and to the large amount of equipment it requires to bring along. Nevertheless, this collectivistic nature characterises the practice widely, leading speleologists to consider it a shared adventure. Whatever the type of trip undertaken (explorative, educational, visiting, rescue), cavers need to coordinate and communicate with each other during their visits. Usually, communication happens through voice, even if it may not be very effective because of environmental noise. However, if during normal trips speleologists bear with the difficulties of voice communication, during rescue operations they overcome the problem by communicating through a phone cable.

**Social adventure**. Speleology is practised in groups: *"as in mountaineering you choose your rope party, and you decide it while desk planning, in caving you form a group because you know that you will complete that exploration only with a specific group"* (I11). However, unlike mountaineering, speleology is necessarily a social sport because *"if you want to go in a cave where you need 5 packs of equipment, then you need 5 people to bring a pack each"* (I03). The need to reach a common goal makes the feeling of living a shared adventure prevail:

> "*The gratification of arriving in a place where no one else has been before is individual at first, of the first person to get in, but just after it is of the group because it is the group that has worked to find a prosecution in the cave. This is fantastic because it creates social unity within the group*" (I09).

> *"What I mean with 'adventure' refers to exploring for exceeding your limits only to a small extent,* [it refers specifically to] *a group, a shared adventure"* (I11).

Furthermore, the explorative team might also include non-cavers covering roles of support, e.g., people looking after the camp and the food:

> *"In speleology, if you want to reach the top, which means going in-depth and explore for miles, you need lots of people. It could be that the base camp is far away, and you need people to bring up all the equipment or to prepare food for you when you get out from a cave after two days of being inside. In the end, you need non-cavers too, but they need to be people who understand how important the team is to achieve the top of the beauty of exploration in speleology"* (I08).

For cavers, communication is fundamental to coordinate inside a cave. As caves are dark and each caver moves at his/her pace, hearing is the only sense that allows them to know where the others are and to coordinate during the progression. Nevertheless, underground voice communication may be difficult: echo and environmental noise being the main obstacles.

> "[Communication] *is hard because there is a terrible echo and often there is water pouring as well [...]. The most important things to hear are 'free!* [When the person behind you can start using the rope]*', 'rock!' if something falls, then the rest is extra"* (I13).

The possibility to use walkie talkies came up in the conversation with some of the interviewees: they were open to their use but sceptical about their effectiveness for the difficult signal propagation that the cave environment offers.

> *"For what concerns normal cave trips, the only possible communication is through voice. The habit to use walkie-talkies that has spread in mountaineering could also be adopted in caving, but their use would be more suitable in large caves, since in narrow ones voice communication is quicker"* (I03).

Even though the environmental noise may prevent speleologists from receiving understandable messages, it appears that the current technological landscape is not able to satisfy the speleologists' needs, who cannot rely on tools commonly used in other outdoor adventure sports.

**Rescue operations.** When planning a field trip, the team creation should consider how many people and what level of expertise is needed to tackle the expected difficulties of the cave they want to visit. The team composition is important because once in the cave the group should be self-sufficient: "*Once the group is formed and enters the cave, no other can join and add resources or skills to the group, meaning that the group has to rely on its members only*" (I12). Safety rules require cave trips to be done in at least four people so that, in case one gets hurt, a partner can stay with the wounded, and the other two can exit the cave and call the rescue service (together, in order to avoid further accidents).

During rescue procedures, communication takes place through a phone cable (twisted pair). This simple technology is the only one used by the groups we interviewed. It is considered the most reliable and flexible method because it allows connection in every spot of the cave without entailing the instability of the Wi-Fi signal. To speak with the others, a rescuer only needs to break the insulating layer of the twisted pair at any point and connect a phone. The rescue service consists of several teams with different roles: one is in charge of ensuring communication between the teams inside the cave and the doctor and the general management outside of it. Their job is to unwind the spools of phone cable and secure it where it cannot hamper other rescuers' passage nor be broken by accident.

> *"In a rescue operation, communication is the second most important thing. First, first aid to the wounded, i.e., thermal blanket, warm food… immediately after, communication"* (I08).

> *"It is a very simple, single pair, copper cable with two poles. We have self-powered handsets, which look like the military ones. Almost every team has a receiver and two clips. They break the insulating layer, connect the clips, and can talk. Obviously, everybody speaks with everybody; there are no phone numbers... we have codes with the ringing, for example, 3 rings to reach team 3, 2 rings - team 1, 1 long ring - the stretcher, many short rings to reach the outside"* (I05).

As the quotation above shows, every team has a receiver to connect to the twisted pair, and they use a different combination of rings to reach the right target of the call. This ring code to identify addressees is not standardised, so at the beginning of each rescue operation it is necessary to review and share it, especially in major operations where teams from different regions or nations collaborate.

> *"Many things are standardised; many others are not… So, what we usually do is that at the beginning* [before starting the operations], *we find an agreement* [on the ring code] *to avoid misunderstandings"* (I13).

Although it is possible to break the twisted pair at any point, it is worth noticing that once the receiver is connected, it is fixed to that spot, and this does not ease the alternation of work for rescue operations and communication with the others:

> *"Every time you arrive at a spot where you will work, the first thing to do is to connect the phone and leave it there. So, you are reachable. You start to work, your partner goes to check things further on, and you get a call. And so, shit! I need to finish what I was doing, go to the phone, answer."* (I05).

Sometimes, the phone cable is used also during the exploration of deep caves (i.e., -1000 m), which last several days and require speleologists to sleep in the cave. In those situations, the twisted pair is rolled out in the cave at the beginning of the expedition and removed at the end of it. In such a context, even if the communication might be disturbed, the phone cable still fulfils the task of "*showing the way to follow to whom come next, as an Ariadne thread*" (I09).

As we may see, speleologists rely on 'old' technological tools for communicating during rescue operations. These appear to be tightly intertwined with the speleologists' routines. Although also the phone cable can be unsatisfying from the point of communication clarity and practicality of use, it is still considered the most reliable technological instrument, being wired in protocols and habits that shape the practice itself. However, speleologists are not completely reluctant to technological advancements even when related to the delicate activity of rescuing, being open to novel devices capable of enriching the communication between 'inside' and 'outside'. Currently, the Italian National Body for the Alpine and Speleological Rescue is considering using video calls to provide physicians outside the cave with a visible and clear image of the wounded:

> *"I have seen that they have found a way to do a video call from inside to the outside by bringing wi-fi in the cave through the phone cable. For the stabilisation of the wounded, it would be very important to transmit vital signals, the conditions, and all the problems outside the cave in real time"(I12).*

*4.3.4. Unearthing: sharing discoveries with others*

The vocation for scientific research is always present in speleology: "*The purpose of every speleology group is to generate new knowledge through new research*" (I12); "*A chamber with stones and some sand… water has brought that sand… when? How does the water work? Caving is a constant study*" (I09). Speleologists' contribution to scientific research occurs mainly through two activities: surveys of newly explored caves and data collection for higher institutions. However, this vocation brings two issues: i) how to train new speleologists, and ii) which results to disseminate and how.

**Interdisciplinarity brings different roles.** Speleology groups are always looking for new young recruits because the exploration is physically demanding and requires much free time: "*the course goal is to 'swell the ranks' of the speleo club. Because we are limited in time, and we don't want the group to disappear*" (I04). During the courses, trainees are primarily taught about the progression in the cave and how to use the gear properly. However, there are also theoretical lessons dedicated to geology, speleogenesis, cave biology, underground topography / meteorology / photography, first aid, and the related equipment. Even if newcomers are not expected to experience all these aspects, this broad introduction has the goal to make them acquire a caver's perspective on the territory, get passionate about searching new caves, and assimilate the scientific vocation of the practice. Then, with time, they develop specific personal interests and their role in the group. Being the practice so diverse, it allows people to find a role that matches with their passions: for example, people more physically prepared will be strong explorers, those with a passion for photography will find great challenges in the darkness of the underground, while those with an interest in technology can learn how to use new tools.

> *"There is this push to document, to study, then everybody has his/her own characteristics at work and in life. Speleologists are not all necessarily professional geologists or biologists, but in our small way, with the* [support of the] *group and with an individual activity of documentation and study, one of us can come to the conclusion that a cave has certain characteristics from a geographical and geological point of view"* (I07).

Over time, the role of older speleologists becomes that of experts, custodians of knowledge, and their main tasks become maintaining relationships with the network (i.e., other groups or political and academic institutions), supporting explorative camps, and disseminating discoveries. By doing so, they can keep contributing to speleology even though they no longer actively explore caves.

**Official reports for institutional audiences.** Thanks to their explorative skills, speleologists are often asked by academic researchers or local institutions to provide support in data collections. For this social and scientific endeavour, they define themselves as 'science labourers'. During the courses we observed, the experienced speleologists mentioned that they contributed to monitoring a stream that provides water to the nearby villages by putting non-polluting markers in the water inside a cave and tracking them down in the rivers and lakes outside. Similarly, they were asked by the local science museum to help to collect data from a glacier by rigging a safe pathway inside the crevasses. Finally, they contribute to survey bat species for Nature 2020[4], a European project for Nature preservation.

As for their own discoveries, to share their findings with other speleologists, academics, and local institutions, speleologists need to document them first. Documentation mainly implies taking note of the GPS coordinates of the cave entrance and tracing the cave map: these data need then to be sent to the regional office responsible for the registration of natural cavities. There are three kinds of maps: i) a plan, i.e., a view that slices the cave horizontally; ii) a profile, i.e., a view that slices the cave lengthwise along the vertical axis; and iii) a cross-section, which is "*another vertical slice of the cavern, this time perpendicular to the explorer's traverse*" (Pérez, 2013, p.302). Maps are created by tracing the plotline through surveying the cave and then drawing the outlines once home. Currently, survey maps are done either in an old-fashioned way, with measuring tape, compass, inclinometer, and a notebook where to write the data; or in a more technological way, by using a laser rangefinder called Disto$^{TM}$ X and the related mobile app TopoDroid. It is worth noticing that this is another example of technology appropriation from other practices and self-build technology. Indeed, Disto$^{TM}$ X is a device usually used by architects to measure rooms, while TopoDroid is an app developed by a speleologist (not from the groups we interviewed) who is a software developer. Furthermore, this technology shows the collaboration that can be established between different groups both at national and international levels:

> "*There is a laser rangefinder called Disto$^{TM}$ X, usually used by architects. I contacted a guy from the Czech Republic to buy the modified version and a British speleo group to buy demagnetised batteries*. [We needed those batteries because the normal ones] *interfere with the compass, and they are not easy to find on the market because they can also be used to build bombs. Then XXX put things together with Bluetooth and the compass as well*" (I12).

Participants enthusiastically welcomed survey technology as it is a mobile and precise tool that speeds up mapping times considerably. New experiments using drones were reported too. However, the enthusiasm showed by the interviewees was jeopardised by the costs and time needed to learn how to use the technology:

> "*Somebody is trying to use drones as well. The drone, protected by a plastic cage, is guided through the cave, and it creates the survey* […]. [Technology] *is a positive thing, but it is expensive* […] *and needs to be learned. I was supposed to learn how to use the Disto$^{TM}$ X in a snap, but I am still studying. You need to dedicate time to new technologies, time and will*" (I04).

To share their work with other speleologists, regional or national meetings are organised every year. Furthermore, all Italian groups can contribute to a national blog[5]. These sharing occasions have the twofold purpose of presenting novel findings (discoveries, survey tools and techniques, collaborations, etc.) and developing a sense of belonging to the wider community.

**Experiential reports for informal audiences.** Speleologists also write informal and experiential reports that they enjoy sharing with other speleologists, trainees, or large audiences to advertise their feats. These experiential reports include not only the data from their discoveries but also the whole explorative experience. Usually, they come in the form of field diaries, photos, and videos. In this regard, we noticed that the evolution of communication means for recording memories creates friction among the different generations of speleologists. As reflected by the quotes of I03 and I09, while the practice of field diary is disappearing, visual reports – especially in the form of videos - are taking over, with the consequent add of certain details to the detriment of others:

> *"It is a lost battle… we cannot make the new generation write things down properly. They just use WhatsApp, not even emails! They do not understand that if we do not keep track of what we do, we will lose memory. This may cause management problems in the future, for example in case we need to report to a funding sponsor like the city hall. They have an aleatory approach. Last year, I got angry because they did an explorative camp and they did not write a single page of diary. There were not many results, but it could have kept trace at least of the activities and participants. We have diaries from the '80s or late '70s that they enjoy reading, but they do not understand, and I feel sorry for that" (I03).*

> *"At the moment, on our website, there are two videos: one is mine, I did it when we explored the so-called 'Penguin branch', and the other one is an I08's video, which he did during the explorative camp of 2016. That's a very nice video that covers all the themes: how we prepared the camp, camp life, what and where we ate, and then it goes in the cave, and there is something on the exploration as well. Conversely, in the Penguin branch video, you can see the first time I entered a new space. […] I noticed another lateral branch, and in the video, you can see me digging in the rocks, then I entered this branch as large as a coffin, but I never reached the end. Maybe it can even be a boring video, but it is useful to see a bit of exploration and to document what we found" (I09).*

---

[4] https://ec.europa.eu/environment/nature/natura2000/index_en.htm (last checked January 2021)
[5] Scintilena, http://www.scintilena.com/#sthash.0WgRJp6W.dpbs (last checked January 2021)

Apart from group memories, the main reason for documenting the experience is the dissemination to the large public. In particular, speleologists find it very hard to explain their underground activities and long for tools capable of recreating their first-hand experiences and the related sensations. For this purpose, one of our participants envisioned an opportunity for Virtual Reality to convey the visual and situated experience of the cave.

> *"Once I saw an app realised by a French group. It was a sort of Google Street view for caves. Practically, you would go on a website and you could see that you were in the wood, at the entrance of a cave. By clicking with the mouse, you could look around and enter the cave. Once in the cave, I could stop and watch around again* […]. *It was* [a] *very short* [path]*, but also very realistic and well done. I remember showing it to a person who had never been in a cave, and s/he got excited. Maybe speleology could benefit from virtual reality to show how caves are inside" (I03).*

## 5. Discussion

So far, HCI research related to the wild natural environment has mostly presented studies about either outdoor sports or citizen science, considering them two separate practices pertaining to two different domains. Conversely, in this paper, we explored speleology, a case study that merges these two domains. By looking at speleology through the theoretical lens of peripheral practices, i.e., "niche, unusual, marginalised and/or highly specialised communities of practice" (Tanenbaum and Tanenbaum, 2018, p.11), we defamiliarized the practice and identified a new perspective on outdoor human activities and the technology that may support them.

Like all practices, speleology consists of physical and mental activities, human bodies, material environments, artefacts, contexts, human capabilities, affinities, and motivations (Kuutti and Bannon, 2014). In this study, we explored the routines and technological artefacts entangled in the speleological practice by analysing its social, embodied, and contextual aspects. We found that speleology combines the characteristics of outdoor adventure sports and citizen science since speleologists' physical performance is needed to discover new caves and then disseminate new knowledge about them. We also showed that speleology is a collaborative practice, in which persons with different roles cooperate to accomplish difficult tasks that pertain to both its adventurous and scientific nature: for instance, exploration implies the collaboration between young and experienced speleologists, and surveying and dissemination - while closely related - entail diversified competencies and duties.

The twofold nature of speleology entails a complex and articulated vision of technology. In fact, speleologists are open to technology that enhances their pursuit of scientific goals, while they are reluctant to artefacts that facilitate the performance or risk to make speleology excessively popular among practitioners who are not sufficiently skilled. The first type of technology includes devices used to 'read' the environment, of which speleologists make extensive use. Moreover, being speleology a niche practice, often cavers do not find suitable devices for their scientific purposes on the market. Therefore, they appropriate technology from other domains or design and self-build novel devices that could support their activities, e.g., by adopting avalanche beacons, range finders, endoscope cameras, or developing the app TopoDroid. Instead, the second group of technology includes devices for the body, for which the reluctance of speleologists reflects the feelings of exclusivity and privilege typical of outdoor adventure sports (Cheverst et al., 2020; Mencarini et al., 2019) and of amateur experts in Citizen Science (Moran et al., 2014), who rely on the expertise gained over the years. By observing speleologists' attitude towards technology through the lens of the 'mediation relations' (Ihde, 1990; Verbeek, 2015), we might notice that technologies instantiating a 'hermeneutic' relation (that is, technologies allowing the 'reading' of the world in new ways) are generally well accepted; whereas technologies instantiating an 'embodiment' relation (that is, technologies that may extend the possibilities of action in the world) might be less accepted, since they may endanger the speleologists' sense of privilege and exclusivity by facilitating the access to the practice.

Our study makes three main contributions to HCI. First, it brings a new perspective on the way people experience the wild outdoors by showing that the physical, technical, and cognitive challenges of outdoor adventure sports can be combined with the pursuit of scientific knowledge. In so doing, it explains the role that technology could play in speleology, where the needs and values of both outdoor adventure sports and citizen science should be addressed. This new perspective may shed light on other practices that put together the efforts of the sports performance and the scientific goal to gain new knowledge, such as snow science, i.e., the survey of how the snowpack changes over time; scientific diving; wildlife sightings, etc. Secondly, this work defines a design space for speleology, i.e., it suggests implications for designing technology specifically addressed to speleology. Finally, it identifies what speleology can do for citizen science, i.e., what can be learned from the self-organised community of speleologists for supporting top-down citizen science projects. In the next subsections, we discuss these two latter contributions.

**5.1. A design space for speleology**

Our study has shown that the speleological practice can be explained through four themes, which correspond to the four core activities performed in speleology: exploration; self- and contextual awareness, group coordination in normal conditions and during rescue operations; documentation, dissemination, and memory (for a summary see Table 3). Currently, for these core activities speleologists use either appropriated or improvised technology and express the desire for some technological support. Here, we provide a series of design implications clustered around the core elements of speleology practice, i.e., the environment, the self, and the group.

*Technology for the environment*

The environment in speleology is both the physical context where the practice takes place and its main goal. Since it is wild, unknown, and hostile, the environment is something to explore and at the same time to tame. Therefore, technology for keeping it under control, foreseeing what comes next, and documenting it is needed.

*Matching means and scope.* Since no part of the cave can be documented without being physically passed through, it would be useful for speleologists to have a technology that combines the tracking of their physical path with the surveying of the cave. The interviewees proposed a sort of Strava that could trace at least the cave plotline by measuring the direction and the distance of their path. As far as we know, currently, there is no technology for tracking physical performance that would be able to also report a measurable account of the environment. The closest device prototyped for a similar function is the underwater camera for divers designed by Hirose et al. (2015): although this camera allows divers to record themselves and the environment more for an experiential report than for a scientific one, its concept could be transferred to the cave setting.

*Knowing what comes next.* At the base of speleology, there is the enjoyment of exploring unknown dark places. Yet, the effort required to discover such places is so high that speleologists cannot afford too many unsuccessful attempts. To reduce the waste of energies and time, speleologists have appropriated technologies from other fields that help them probe the ground. This shows their need for a technology that supports them foresee where caves are and what comes next underground, like a probe or a radar.

*Keeping a connection with the outside.* It would be important to provide cavers with the awareness of the passing of time and the change of weather conditions outside the cave so to allow them to adjust their plans accordingly. For example, in case of sudden storms, cavers could expect floods inside the cave and react promptly thereupon, while in case of snow, they could save energies to keep attention higher than usual when leaving the cave. Since speleologists always inform other people about their trips or have supporting partners at the base camp, communication with these external reference points would help avoid accidents. For this purpose, technology should enable speleologists to keep always a connection with the outside.

*Technology for the self*

Individuals in speleology are always part of a group. Technology for the self would serve to ensure speleologists to feel at ease inside a cave both physically and mentally, so that the whole group can pursue its goal and perform its activity safely.

*Self-awareness.* The preservation of sufficient energies to go back to the surface is extremely important to avoid accidents or putting the group in difficulty. A device able to measure a caver's physical conditions in terms of blood oxygenation, body temperature, calories, etc., and to show this information during the entire cave trip would be useful to prevent physical or mental crises. An example of such technology could be the 'social fabric fitness' displays proposed by Mauriello et al. (2014), which would work well in the darkness of a cave, because they are visual, inform both the wearer and the other group members, and serve primarily for the ongoing practice rather than for self-reflection afterwards.

*Following the traces.* Expressions used by our participants like 'the rope is the road' and 'the phone cable as Ariadne's thread' make us think that, even if orientation and the possibility to get lost inside a cave have never been reported as a big problem, seeing traces from the passage of others is reassuring. Therefore, it could be useful to have visual feedback about the passage or the activity of others. Furthermore, if the traces are physical devices left in place, they could help cavers communicate with the outside (if properly equipped with a short-range transmission apparatus, such as a Zigbee mesh network).

### *Technology for the group*
Speleology is a collective practice. Both inside and outside the cave, speleologists are part of a group and a wider community. Technology could help speleologists communicate during cave explorations with their team members and, later on, share their discoveries with others.

*Communication.* Although in normal conditions communication has not been reported to be a major issue, overcoming environmental noise would be beneficial for coordination during standard progressions. If walkie-talkies are not reliable in the context of caves, we might think of wearable devices for exchanging simple visual messages, like the 'glanceable displays' (Gouveia et al., 2016). These displays offer an intuitive visual representation of the data collected by a wearable device and require just a glance to interact with them. As for the context of speleology, these displays could encode the few standard commands that speleologists exchange to coordinate themselves into the visual messages that team members could intentionally send to each other. This feature might be integrated into the wearable devices for self-awareness of following the others' traces to exploit the short-range mesh network.

*Saving lives.* Concerning rescue operations, during which often the physician remains outside the cave, interviewees told us that they were already considering using video calls to show the wounded to the doctor. Video calls would be more informative if the injured person's biosignals could be integrated.

*Sharing the privilege.* Even if speleologists would allow only other trained speleologists to access caves, they long for sharing their privileged experience with wider audiences. Immersive technology such as 3D virtual environments could provide an accurate reproduction of a first-hand caving experience. This kind of technology would be useful for disseminating the experience to the general population, as well as to expert and peer audiences. In this regard, a desirable evolution could be creating comprehensive records of experiences, where immersive visual representations are enriched with historical and scientific data.

### **5.2. Speleology for Citizen Science**
Speleology (as we have encountered it) is carried out by autonomous groups of people who pursue the discovery of new caves. However, we believe that speleology can be considered a self-organized form of citizen science because it is practised by amateur experts and offers a concrete example of autonomous management of the scientific activity, potentially providing insights on how to address the issues concerning citizen science projects.

As we have seen in the Related Work section, these issues are related to i) the different levels of citizens' engagement, which denote specific goals and types of citizen science projects (Aoki et al., 2017; Bonney et al., 2009; Qaurooni et al., 2016); ii) the sustainability of citizen science projects over the long term (Mugar et al., 2014; Rotman et al., 2012; Tinati et al., 2015); iii) the quality of data with respect to the level of citizens' expertise (Elbroch et al., 2011); iv) and the design of technology to support the activity (Cottman-Fields et al., 2013; Gaver et al., 2019; Phillips et al., 2014; Polys et al., 2020). While the latter point has already been addressed in Section 5.1, in the following, we summarise how speleologists address the former three challenges highlighting several insights for citizen science.

**Independency of goals**. Regarding the different levels of engagement in the practice, the results emerged from our study show that speleologists are highly engaged and intrinsically motivated because they are expert amateurs with extensive local knowledge, that most of the time set the goal of their research autonomously, e.g., when they decide which cave to explore. However, at times, they engage in activities organized by higher institutions (e.g., when the science museum asked them to do measurements in the depth of a glacier): on these occasions, they assume the role of 'science labourers'. In this respect, designs favouring participation, choice, and deliberation about goals and methods to be used in citizen science projects would increase the citizens' sense of agency, belonging, and commitment to their communities.

**A community**. The quality of data in speleology is ensured by being part of a community at multiple levels. First, speleology is made by groups of volunteers whose ranks are enriched through the annual organisation of courses. Both practical and theoretical training of newcomers is carried out by long-serving practitioners, who are experts in the subdisciplines of speleology. Second, at a broader level, speleologists are part of a network of groups that organise annual congresses (either regional or national) where both speleologists and academics participate. These meetings motivate them to stick to their mission and commit to making new discoveries as well as discuss methods and tools with peers and collaborate with academics to analyse the collected materials. From this perspective, citizen science projects could give more room for learning activities aimed at developing fundamental skills and knowledge in newcomers. 'Expert citizens' could then be paired with novices and act as mentors, that is, guiding them. To this aim, citizen science communities could leverage techniques employed in social matching systems, which create and match user profiles based on interests and competencies, recommending people to each other (Terveen and McDonald, 2005). Moreover, citizen science projects could give more visibility to their outcomes even outside the strict circle of their community, encouraging their members to share data and results with other similar citizen science communities or platforms and to organize online events open to the 'outside world.'

**A mix of different competences and roles.** The sustainability of speleology groups in the long term is reached through differentiation of roles and tasks. Being a discipline with many sub-activities, participants with different interests and different physical abilities can find different roles in the micro-society of a speleology group. For example, the inevitable decrease in personal strength corresponds to the accumulation of experience and an increment of theoretical knowledge. So, if the youth are more gifted for physical performance and exploration thanks to their physical skills, elders become the experts, providing suggestions and maintaining relationships within the wider network, i.e., with political and academic institutions, other groups, etc. In this way, speleologists belonging to the same group collaborate, each one contributing in a different manner and extent according to his/her own interests and capabilities. Likewise, citizen science projects could develop organisational structures relying on the differentiation of roles, rather than standardisation, as well as design 'paths' in which new roles are progressively made available as long as the citizens' permanence in the community increases. The opportunity to progressively access new roles that better match with the current citizen's interests, skills, and availability could represent a supplementary motivational drive for remaining engaged in the community as demonstrated by Aoki et al. (2017) and Rotman et al. (2012).

Given the productivity and long history of the groups we interviewed, we deem them a good example to inspire successful and sustainable citizen science projects. Citizen science projects proposed by research institutions that need to collect large amounts of data should consider the organisational characteristics of autonomous speleology groups in order to keep participants engaged over time. To this end, key aspects are allowing participants to be autonomous in setting their goals and taking responsibility for their achievements, cover different roles with different levels of engagement according to their interests and competencies, build a network of relations within both the working group and the wider network in order to exchange results and methods.

## 6. Limitations and future works

This work presents some limitations. First, it was conducted in one nation and specifically in one mountain region. Although our findings resonate with those of Pérez in Venezuela (2015) and Cant's studies in the United Kingdom (2003, 2006), future work could try drawing from various sources at different latitudes to produce more generalizable insights. Second, many aspects of the practice presented in this study could be further explored. For instance, we found some evidence that speleology is a highly intrinsically motivated practice, as speleologists' self-organization and perseveration, despite the difficulties that they have to face during the exploration, may show. Nevertheless, we did not investigate motivational factors in depth since our focus mainly revolved around the routines and the embodied activities entangled with the practice. Future work could explore speleology from a psychological perspective to unfold all the motivational and attitudinal aspects involved. Similarly, a 'social and organizational lens' focusing on the functioning of a collaborative group of speleologists could further shed light, for example, on the leadership dynamics, the conflicts that might arise within the group, what structure and roles might exist and how these might change dynamically over time. Finally, another interesting direction for future work may include the investigation of other sporting sciences, i.e., other practices that merge aspects of outdoor adventure sports and science, such as snow science and scientific diving, to trace differences and common traits with our findings.

## 7. Conclusions

Over recent years, HCI has investigated human activities related to Nature, such as outdoor sports, adventure, and citizen science. Nevertheless, HCI has treated each of these domains separately: until now, no study has investigated hybrid ways to experience the outdoors that merge the pleasure of physical performance with that of generating new knowledge. In this article, we addressed this knowledge gap by offering an understanding of the peripheral yet multifaceted human practice of speleology, a leisure activity practised by people united by the willingness to discover new underground places and the ownership of specific physical skills. We showed that speleologists have an ambivalent attitude towards technology: on the one hand, speleology welcomes technological artefacts supporting a scientific endeavour; on the other hand, it resists forms of technological facilitation that may weaken their challenge against Nature. Finally, this study provides suggestions both for the design of devices supporting speleology and for the organisation of sustainable long-term citizen science projects.

## Acknowledgements

We thank all the speleologists from the GSL and the GST groups in Trento (Italy) who voluntarily took part in the interviews. The first author gratefully acknowledges the grant from Fondazione Cassa di Risparmio di Trento e Rovereto (grant n. 1939/17).

## References

Adcock, M., Anderson, S., Berkvosky, S., Flick, P., Frousheger, D., Grandbois, B., Gunn, C., Haddon, D., Li, J., Lowe, T., Mackey, B., Pauling, F., Richter, C., Stepanas, K., Talbot, F., Walker, G., Ward, B., Tomes, L., Miles, S., Keogh, D., Colenso, B., Hay, W.L., Longmore, D., Hanly, J., Canavan, 2015. Exploring the Jenolan Caves: bringing the physical world to 3D online education, in: Proceedings of the 20th International Conference on 3D Web Technology. p. 161.

Ahtinen, A., Isomursu, M., Huhtala, Y., Kaasinen, J., Salminen, J., Häkkilä, J., 2008. Tracking Outdoor Sports – User Experience Perspective, in: European Conference on Ambient Intelligence. Springer, pp. 192–209. https://doi.org/10.1007/978-3-540-89617-3_13

Anderson, Z., Lusk, C., Jones, M.D., 2017. Towards understanding hikers' technology preferences, in: Proceedings of the 2017 ACM International Joint Conference on Pervasive and Ubiquitous Computing and Proceedings of the 2017 ACM International Symposium on Wearable Computers. ACM Press, pp. 1–4. https://doi.org/10.1145/3123024.3123089

Aoki, P., Woodruff, A., Yellapragada, B., Willett, W., 2017. Environmental protection and agency: Motivations, capacity, and goals in participatory sensing, in: Proceedings of the 2017 CHI Conference on Human Factors in Computing Systems. pp. 3138–3150.

Beraldin, J.-A., Blais, F., Cournoyer, L., Picard, M., Gamache, D., Valzano, V., Bandiera, A., Gorgoglione, M., 2006. Multi-resolution digital 3D imaging system applied to the recording of grotto sites: the case of the Grotta dei Cervi, in: Proceedings of the 7th International Conference on Virtual Reality, Archaeology and Intelligent Cultural Heritage. Eurographics Association, pp. 45–52.

Bidwell, N.J., Browning, D., 2010. Pursuing genius loci: interaction design and natural places. Personal and Ubiquitous Computing 14, 15–30.

Bonney, R., Ballard, H., Jordan, R., McCallie, E., Phillips, T., Shirk, J., Wilderman, C.C., 2009. Public Participation in Scientific Research: Defining the Field and Assessing Its Potential for Informal Science Education. A CAISE Inquiry Group Report. Online Submission.

Braun, V., Clarke, V., 2006. Using thematic analysis in psychology. Qualitative Research in Psychology 3, 77–101. https://doi.org/10.1191/1478088706qp063oa

Cant, S.G., 2006. British speleologies: geographies of science, personality and practice, 1935–1953. Journal of Historical Geography 32, 775–795.

Cant, S.G., 2003. The Tug of Danger with the Magnetism of Mystery' Descents into 'the Comprehensive, Poetic-Sensuous Appeal of Caves. Tourist Studies 3, 67–81.

Cheverst, K., Bødker, M., Daiber, F., 2020. Design Sensitivities for Technology in Mountaineering, in: HCI Outdoors: Theory, Design, Methods and Applications, Human-Computer Interaction Series. Springer, Cham, pp. 197–211.

Cottman-Fields, M., Brereton, M., Roe, P., 2013. Virtual birding: extending an environmental pastime into the virtual world for citizen science, in: Proceedings of the SIGCHI Conference on Human Factors in Computing Systems. pp. 2029–2032.

Daiber, F., Kosmalla, F., Wiehr, F., Krüger, A., 2017. Follow the pioneers: towards personalized crowd-sourced route generation for mountaineers, in: Proceedings of the 2017 ACM International Joint Conference on Pervasive and Ubiquitous Computing and Proceedings of the 2017 ACM International Symposium on Wearable Computers. ACM, Maui Hawaii, pp. 1051–1055. https://doi.org/10.1145/3123024.3124447

Daiber, F., Kosmalla, F., Wiehr, F., Krüger, A., 2016. Outdoor Nature Lovers vs. Indoor Training Enthusiasts: A Survey of Technology Acceptance of Climbers, in: Proceedings of the SIGCHI Conference on Human Factors in Computing Systems - CHI'16.

Davidson, L., Stebbins, R.A., 2011. Serious leisure and nature: sustainable consumption in the outdoors. Springer.

Desjardins, A., Neustaedter, C., Greenberg, S., Wakkary, R., 2014. Collaboration Surrounding Beacon Use During Companion Avalanche Rescue, in: Proceedings of the 17th ACM Conference on Computer Supported Cooperative Work & Social Computing - CSCW'14. pp. 877–887. https://doi.org/10.1145/2531602.2531684

Elbroch, M., Mwampamba, T.H., Santos, M.J., Zylberberg, M., Liebenberg, L., Minye, J., Mosser, C., Reddy, E., 2011. The value, limitations, and challenges of employing local experts in conservation research. Conservation Biology 25, 1195–1202.

Fails, J.A., Herbert, K.G., Hill, E., Loeschorm, C., Kordecki, S., Dymko, D., DeStefano, A., Zill, C., 2014. GeoTagger: a collaborative and participatory environmental inquiry system, in: Proceedings of the Companion Publication of the 17th ACM Conference on Computer Supported Cooperative Work & Social Computing. Presented at the CSCW, pp. 157–160.

Gaver, W., Boucher, A., Vanis, M., Sheen, A., Brown, D., Ovalle, L., Matsuda, N., Abbas-Nazari, A., Phillips, R., 2019. My Naturewatch Camera: Disseminating Practice Research with a Cheap and Easy DIY Design, in: Proceedings of the 2019 CHI Conference on Human Factors in Computing Systems. pp. 1–13.

Gouveia, R., Pereira, F., Karapanos, E., Munson, S.A., Hassenzahl, M., 2016. Exploring the design space of glanceable feedback for physical activity trackers, in: Proceedings of the 2016 ACM International Joint Conference on Pervasive and Ubiquitous Computing. pp. 144–155.

Häkkilä, J., Bidwell, N.J., Cheverst, K., Colley, A., Kosmalla, F., Robinson, S., Schöning, J., 2018. Reflections on the NatureCHI workshop series: unobtrusive user experiences with technology in nature. International Journal of Mobile Human Computer Interaction (IJMHCI) 10, 1–9.

Hirose, M., Sugiura, Y., Minamizawa, K., Inami, M., 2015. PukuPuCam: A recording system from third-person view in scuba diving, in: Proceedings of the 6th Augmented Human International Conference. pp. 161–162.

Hsu, Y.-C., Dille, P., Cross, J., Dias, B., Sargent, R., Nourbakhsh, I., 2017. Community-empowered air quality monitoring system, in: Proceedings of the 2017 CHI Conference on Human Factors in Computing Systems. pp. 1607–1619.

Ihde, D., 1990. Technology and the lifeworld: From garden to earth, The Indiana series in the philosophy of technology.

Kuutti, K., Bannon, L.J., 2014. The turn to practice in HCI, in: Proceedings of the 32nd Annual ACM Conference on Human Factors in Computing Systems - CHI '14. pp. 3543–3552. https://doi.org/10.1145/2556288.2557111

Mattes, J., 2015. Underground fieldwork – A cultural and social history of cave cartography and surveying instruments in the 19th and at the beginning of the 20th century. International Journal of Speleology 44, 251–266.

Mauriello, M., Gubbels, M., Froehlich, J.E., 2014. Social Fabric Fitness: The Design and Evaluation of Wearable E-Textile Displays to Support Group Running, in: Proceedings of the SIGCHI Conference on Human Factors in Computing Systems - CHI'14. pp. 2833–2842. https://doi.org/10.1145/2556288.2557299

McCrickard, S.D., Jones, M., Stelter, T.L. (Eds.), 2020. HCI Outdoors: Theory, Design, Methods and Applications, Human-Computer Interaction series. Springer, Cham.

McDonald, N., Schoemebeck, S., Forte, A., 2019. Reliability and Inter-rater Reliability in Qualitative Research: Norms and Guidelines for CSCW and HCI Practice. CSCW, Proceedings, of the ACM Human-Computer Interaction 3, 1–23. https://doi.org/doi.org/10.1145/3359174

Mencarini, E., Leonardi, C., Cappelletti, A., Giovanelli, D., De Angeli, A., Zancanaro, M., 2019. Co-designing wearable devices for sports: The case study of sport climbing. International Journal of Human-Computer Studies 124, 26–43.

Moran, S., Pantidi, N., Rodden, T., Chamberlain, A., Griffiths, C., Zilli, D., Merrett, G., Rogers, A., 2014. Listening to the forest and its curators: lessons learnt from a bioacoustic smartphone application deployment, in: Proceedings of the SIGCHI Conference on Human Factors in Computing Systems. pp. 2387–2396.

Mugar, G., Østerlund, C., Hassman, K.D., Crowston, K., Jackson, C.B., 2014. Planet hunters and seafloor explorers: legitimate peripheral participation through practice proxies in online citizen science, in: Proceedings of the 17th ACM Conference on Computer Supported Cooperative Work & Social Computing. pp. 109–119.

Müller, F. "Floyd", Pell, S.J., 2016. Technology meets Adventure: Learnings from an Earthquake-Interrupted Mt. Everest Expedition, in: UbiComp. pp. 817–828.

Olson, J.S., Kellogg, W.A. (Eds.), 2014. Ways of Knowing in HCI. Springer, New York, NY, USA.

Pérez, M.A., 2015. Exploring the vertical: science and sociality in the field among cavers in Venezuela. Social & Cultural Geography 16, 226–247. https://doi.org/10.1080/14649365.2014.973438

Phillips, R.D., Blum, J.M., Brown, M.A., Baurley, S.L., 2014. Testing a grassroots citizen science venture using open design, "the bee lab project," in: CHI'14 Extended Abstracts on Human Factors in Computing Systems. pp. 1951–1956.

Pike, E.C.J., Beames, S.K., 2013. Outdoor Adventure and Social Theory, Sport Studies. Routledge.

Polys, N., Sforza, P., Munsell, J., 2020. PlantShoe: Botanical Detectives, in: HCI Outdoors: Theory, Design, Methods, Applications, Human-Computer Interaction Series. Springer, Cham, pp. 119–136.

Posti, M., Schöning, J., Häkkilä, J., 2014. Unexpected journeys with the HOBBIT: the design and evaluation of an asocial hiking app, in: Proceedings of the 2014 Conference on Designing Interactive Systems. pp. 637–646.

Qaurooni, D., Ghazinejad, A., Kouper, I., Ekbia, H., 2016. Citizens for science and science for citizens: The view from participatory design, in: Proceedings of the 2016 CHI Conference on Human Factors in Computing Systems. pp. 1822–1826.

Rotman, D., Preece, J., Hammock, J., Procita, K., Hansen, D., Parr, C., Lewis, D., Jacobs, D., 2012. Dynamic changes in motivation in collaborative citizen-science projects, in: Proceedings of the ACM 2012 Conference on Computer Supported Cooperative Work. pp. 217–226.

Schuchardt, P., Bowman, D.A., 2007. The benefits of immersion for spatial understanding of complex underground cave systems, in: Proceedings of the 2007 ACM Symposium on Virtual Reality Software and Technology - VRST '07. ACM Press, pp. 121–124. https://doi.org/10.1145/1315184.1315205

Silvestre, I., Rodrigues, J.I., Figueiredo, M., Veiga-Pires, C., 2013. Framework for 3D data modeling and web visualization of underground caves using open source tools, in: Proceedings of the 18th International Conference on 3D Web Technology. pp. 121–128.

Su, N.M., Cheon, E., 2017. Reconsidering nature: The dialectics of fair chase in the practices of American Midwest hunters, in: Proceedings of the SIGCHI Conference on Human Factors in Computing Systems. Presented at the CHI'17, ACM, pp. 6089–6100.

Tanenbaum, J., Tanenbaum, K., 2018. Steampunk, Survivalism and Sex Toys: An Exploration of How and Why HCI Studies Peripheral Practices, in: Filimowicz, M., Tzankova, V. (Eds.), New Directions in Third Wave Human-Computer Interaction: Volume 2 - Methodologies. Springer, Cham, pp. 11–24.

Terveen, L., McDonald, D.W., 2005. Social matching: A framework and research agenda. ACM Transactions on Computer-Human Interaction 12, 401–434. https://doi.org/0.1145/1096737

Tholander, J., Nylander, S., 2015. Snot, Sweat, Pain, Mud, and Snow: Performance and Experience in the Use of Sports Watches, in: Proceedings of the SIGCHI Conference on Human Factors in Computing Systems - CHI'15. pp. 2913–2922. https://doi.org/10.1145/2702123.2702482

Tinati, R., Van Kleek, M., Simperl, E., Luczak-Rösch, M., Simpson, R., Shadbolt, N., 2015. Designing for citizen data analysis: A cross-sectional case study of a multi-domain citizen science platform, in: Proceedings of the 33rd Annual ACM Conference on Human Factors in Computing Systems. pp. 4069–4078.

Verbeek, P.-P., 2015. Beyond interaction: a short introduction to mediation theory. Interactions 22, 26–31.

Wang, J.Z., Ge, W., Snow, D.R., Mitra, P., Giles, C.L., 2010. Determining the sexual identities of prehistoric cave artists using digitized handprints: a machine learning approach, in: Proceedings of the 18th ACM International Conference on Multimedia. pp. 1325–1332.

Wheaton, B., 2004. Understanding lifestyle sports: Consumption, Identity and Difference. Routledge.

Woźniak, P.W., Fedosov, A., Mencarini, E., Knaving, K., 2017. Soil, Rock, and Snow: On Designing for Information Sharing in Outdoor Sports, in: Proceedings of the 2017 Conference on Designing Interactive Systems. ACM, pp. 611–623.